\documentclass[10pt, final, twocolumn, twoside, romanappendices]{IEEEtran}
\usepackage{amsmath,amssymb}
\usepackage[bookmarks,colorlinks]{hyperref} 
\usepackage{acronym}  
\usepackage{color}  
\usepackage{soul}
\usepackage{graphicx,psfrag,subcaption,cite}
\usepackage{mathtools}
\DeclarePairedDelimiter\ceil{\lceil}{\rceil}

\usepackage{bbm}
\usepackage{bigints}
\usepackage{mathrsfs,cancel}
\usepackage[table]{xcolor}
\usepackage{algorithm}
\usepackage[end]{algpseudocode}
\usepackage{arydshln}

\usepackage{tikz}
\usepackage{pgfplotstable}
\usepackage{rotate}

\definecolor{rubblue}{cmyk}{1,0.5,0,0.6}
\definecolor{rubgreen}{cmyk}{0.5,0,1,0}
\definecolor{rubgray}{cmyk}{0.03,0.03,0.03,0.1}

\usepgfplotslibrary{units}

\usetikzlibrary{%
patterns,%
calc,%
fit,%
arrows,%
plotmarks,%
shadows,%
chains,%
shapes%
}

\tikzset{>=latex'} 
\tikzstyle{every picture}+=[remember picture] 
\pgfdeclarelayer{background}
\pgfdeclarelayer{foreground}
\pgfsetlayers{background,main,foreground}

\tikzstyle{blueblock}=[draw=rubblue, rectangle, thick, drop shadow, minimum width=20mm, minimum height=8mm,fill=rubblue!20, text width=20mm, text centered]
\tikzstyle{bluebox}=[draw=rubblue, rectangle, thick, drop shadow, minimum width=8mm, minimum height=8mm,fill=rubblue!20, text width=8mm, text centered]
\tikzstyle{greenblock}=[draw=rubgreen, rectangle, thick, drop shadow, minimum width=20mm, minimum height=8mm,fill=rubgreen!20, text width=20mm, text centered]
\tikzstyle{dot} = [draw, circle, minimum size=0.2pt,scale=0.3,fill=black,black]
\tikzstyle{smalldot} = [draw, circle, minimum size=0.1pt,scale=0.2,fill=black,black]
\tikzstyle{reddot}  =[draw,circle,minimum size=0.2pt,scale=0.8,fill=red,thin]
\tikzstyle{greendot}  =[draw,circle,minimum size=0.2pt,scale=0.8,fill=Green,thin]
\tikzstyle{bluedot}  =[draw,circle,minimum size=0.2pt,scale=0.8,fill=blue,thin]
\tikzstyle{whitedot}=[draw,circle,minimum size=0.2pt,scale=0.8,fill=white,thin]
\tikzstyle{blackdot} = [draw, circle, minimum size=0.2pt,scale=0.7,fill=black,black]
\tikzstyle{sum} = [drop shadow, draw=rubblue, thick, fill=rubblue!20, circle]
\tikzstyle{relay} = [blueblock, minimum width=5mm, minimum height=20mm, text width=5mm, rounded corners=2pt]
\tikzstyle{relay2} = [blueblock, minimum width=5mm, minimum height=15mm, text width=5mm, rounded corners=2pt]
\tikzstyle{relay3} = [blueblock, minimum width=5mm, minimum height=25mm, text width=5mm, rounded corners=2pt]
\tikzstyle{relay4} = [blueblock, minimum width=5mm, minimum height=10mm, text width=5mm, rounded corners=2pt]
\tikzstyle{relay5} = [blueblock, minimum width=5mm, minimum height=50mm, text width=5mm, rounded corners=2pt]
\tikzstyle{relay6} = [blueblock, minimum width=5mm, minimum height=5mm, text width=5mm, rounded corners=2pt]
\tikzstyle{circgreen} = [draw, circle, inner sep=2pt, fill=rubgreen, drop shadow, thick]
\tikzstyle{circwhite} = [draw, circle, inner sep=2pt, fill=white, drop shadow, thick]
\tikzstyle{circdashed} = [draw, dashed, circle, inner sep=2pt, fill=rubgray, drop shadow, thick]
\tikzstyle{vertbox} = [rectangle, draw=rubblue, thick, rotate=90, text centered, minimum width=16.5mm, minimum height=8mm, text width=16.5mm, inner sep=0pt, fill=rubblue!20, drop shadow]
\tikzstyle{vertboxb} = [rectangle, draw=rubblue, thick, rotate=90, text centered, minimum width=16.5mm, minimum height=8mm, text width=16.5mm, fill=rubblue!20, drop shadow]
\tikzstyle{vertboxshort} = [rectangle, draw=rubblue, thick, rotate=90, text centered, minimum width=10mm, minimum height=8mm, text width=10mm, inner sep=0pt, fill=rubblue!20, drop shadow]
\tikzstyle{smalldotgreen} = [draw=rubgreen, circle, minimum size=0.2pt,scale=0.8,fill=rubgreen!20]
\tikzstyle{antenna} = [regular polygon, regular polygon sides=3, draw, shape border rotate=180, minimum size=0.2pt, scale=0.3]

\tikzstyle{poly} = [regular polygon, regular polygon sides=6, shape aspect=0.5, minimum width=1.5cm, minimum height=0.35cm, draw, dashed]

\definecolor{cff9e00}{RGB}{255,158,0}
\definecolor{c4fff00}{RGB}{79,255,0}
\definecolor{cff0012}{RGB}{255,0,18}
\definecolor{c00c5ff}{RGB}{0,197,255}
\definecolor{c046f00}{RGB}{4,111,0}
\definecolor{c004b9d}{RGB}{0,75,157}

\newlength{\mylen}
\usetikzlibrary{petri}
\usetikzlibrary{shapes}
\usetikzlibrary{positioning,arrows,patterns}
\usepackage[utf8x]{inputenc}
\usepackage{scalefnt}
\usetikzlibrary{calc,decorations.markings}
\pgfplotsset{compat=1.10}
\usetikzlibrary{arrows.meta}
\usepackage[columnwise,switch,mathlines]{lineno} 

\usepackage{pgfplots}
\pgfplotsset{compat=newest}
\usepgfplotslibrary{groupplots}

\usepackage{bm}

\DeclareMathAlphabet{\mathsfbr}{OT1}{cmss}{m}{n}
\SetMathAlphabet{\mathsfbr}{bold}{OT1}{cmss}{bx}{n}
\DeclareRobustCommand{\msf}[1]{%
  \ifcat\noexpand#1\relax\msfgreek{#1}\else\mathsfbr{#1}\fi
}

\makeatletter
\newcommand{\msfgreek}[1]{\csname s\expandafter\@gobble\string#1\endcsname}
\makeatother

\DeclareFontEncoding{LGR}{}{} 
\DeclareSymbolFont{sfgreek}{LGR}{cmss}{m}{n}
\SetSymbolFont{sfgreek}{bold}{LGR}{cmss}{bx}{n}
\DeclareMathSymbol{\salpha}{\mathord}{sfgreek}{`a}
\DeclareMathSymbol{\sbeta}{\mathord}{sfgreek}{`b}
\DeclareMathSymbol{\sgamma}{\mathord}{sfgreek}{`g}
\DeclareMathSymbol{\sdelta}{\mathord}{sfgreek}{`d}
\DeclareMathSymbol{\sepsilon}{\mathord}{sfgreek}{`e}
\DeclareMathSymbol{\szeta}{\mathord}{sfgreek}{`z}
\DeclareMathSymbol{\seta}{\mathord}{sfgreek}{`h}
\DeclareMathSymbol{\stheta}{\mathord}{sfgreek}{`j}
\DeclareMathSymbol{\siota}{\mathord}{sfgreek}{`i}
\DeclareMathSymbol{\skappa}{\mathord}{sfgreek}{`k}
\DeclareMathSymbol{\slambda}{\mathord}{sfgreek}{`l}
\DeclareMathSymbol{\smu}{\mathord}{sfgreek}{`m}
\DeclareMathSymbol{\snu}{\mathord}{sfgreek}{`n}
\DeclareMathSymbol{\sxi}{\mathord}{sfgreek}{`x}
\DeclareMathSymbol{\somicron}{\mathord}{sfgreek}{`o}
\DeclareMathSymbol{\spi}{\mathord}{sfgreek}{`p}
\DeclareMathSymbol{\srho}{\mathord}{sfgreek}{`r}
\DeclareMathSymbol{\ssigma}{\mathord}{sfgreek}{`s}
\DeclareMathSymbol{\stau}{\mathord}{sfgreek}{`t}
\DeclareMathSymbol{\supsilon}{\mathord}{sfgreek}{`u}
\DeclareMathSymbol{\sphi}{\mathord}{sfgreek}{`f}
\DeclareMathSymbol{\schi}{\mathord}{sfgreek}{`q}
\DeclareMathSymbol{\spsi}{\mathord}{sfgreek}{`y}
\DeclareMathSymbol{\somega}{\mathord}{sfgreek}{`w}

\DeclareMathSymbol{\svarsigma}{\mathord}{sfgreek}{`c}

\DeclareMathSymbol{\sGamma}{\mathalpha}{sfgreek}{`G}
\DeclareMathSymbol{\sDelta}{\mathalpha}{sfgreek}{`D}
\DeclareMathSymbol{\sTheta}{\mathalpha}{sfgreek}{`J}
\DeclareMathSymbol{\sLambda}{\mathalpha}{sfgreek}{`L}
\DeclareMathSymbol{\sXi}{\mathalpha}{sfgreek}{`X}
\DeclareMathSymbol{\sPi}{\mathalpha}{sfgreek}{`P}
\DeclareMathSymbol{\sSigma}{\mathalpha}{sfgreek}{`S}
\DeclareMathSymbol{\sUpsilon}{\mathalpha}{sfgreek}{`U}
\DeclareMathSymbol{\sPhi}{\mathalpha}{sfgreek}{`F}
\DeclareMathSymbol{\sPsi}{\mathalpha}{sfgreek}{`Y}
\DeclareMathSymbol{\sOmega}{\mathalpha}{sfgreek}{`W}

\DeclareRobustCommand{\mcal}[1]{%
  \ifcat\noexpand#1\relax\mathnormal{#1}\else\cal{#1}\fi
}
\DeclareRobustCommand{\BM}[1]{%
  \ifcat\noexpand#1\relax\bm{\boldUppercaseItalicGreek{#1}}\else\bm{#1}\fi
}
\makeatletter
\newcommand{\boldUppercaseItalicGreek}[1]{\csname var\expandafter\@gobble\string#1\endcsname}
\makeatother
\newcommand{\rv}[1]{\msf{#1}}
\newcommand{\RV}[1]{\bm{\msf{#1}}}

\newcommand{\V}[1]{\bm{#1}}
\newcommand{\M}[1]{\BM{#1}}

\newcommand{\beq}{\begin{equation}}
\newcommand{\eeq}{\end{equation}}

\newtheorem{theorem}{Theorem}

\newtheorem{proposition}{Proposition}

\newtheorem{lemma}{Lemma}

\newtheorem{remark}{Remark}

\definecolor{myblack}{rgb}{0,0,0}
\definecolor{myblack2}{rgb}{0,0,0}
\newcommand{\paperTitle}{Rate Adaptation and Latency in Heterogeneous IoT Networks}
\newcommand{\paperTitleMarkboth}{Rate Adaptation and Latency in Heterogeneous IoT Networks}

\acrodef{IoT}{Internet of things}
\acrodef{TSP}{transmission success probability}
\acrodef{SIR}{signal-to-interference ratio}
\acrodef{MC}{{Markov chain}}

\definecolor{mygreentext}{RGB}{31,97,25} 
\definecolor{myredtext}{RGB}{207,21,24}
\definecolor{mybluetext}{RGB}{0,43,187} 
\definecolor{myblacktext}{RGB}{0,0,0} 
\definecolor{mygray}{RGB}{200,200,200} 
\definecolor{myorange}{RGB}{255, 178, 102}
\definecolor{mymagenta}{rgb}{0.78, 0.08, 0.52}
\definecolor{mymyblacktext}{rgb}{0, 0.74, 1}

\newcommand{\savespaces}{\vspace{-4pt}}

	\algnewcommand{\Initialize}[1]{%
	\State \textbf{Initialize:}
	\Statex \hspace*{\algorithmicindent}\parbox[t]{.8\linewidth}{\raggedright #1}
}
\begin{document}




\title{\paperTitle}


\author{
	\vspace{0.2cm}
       Hesham~ElSawy,~\IEEEmembership{Senior~Member,~IEEE} 
        \thanks{H.\ ElSawy is with the Electrical Engineering Department, King Fahd University of Petroleum \& Minerals (KFUPM), Dhahran, Saudi Arabia, (e-mail: \texttt{hesham.elsawy@kfupm.edu.sa}).}
     \vspace{-0.5cm}
}
\maketitle 
\markboth{ElSawy: \paperTitleMarkboth}{ElSawy: \paperTitleMarkboth}


\setcounter{page}{1}

\begin{abstract}
This paper studies the effect of rate adaptation in time slotted Internet of things (IoT) networks. For a given time slot duration and packets size, rate adaptation necessitates packet fragmentation to fit the time slot duration. Accounting for the quality and time resolution of the underlying traffic, this paper characterizes the tradeoff between transmission rate and packet latency in IoT networks. Using tools from stochastic geometry and queueing theory, a novel mathematical framework is developed for static and dynamic rate adaptation schemes. The results show that there is an optimal static rate that minimizes latency, which depends on the network parameters. Furthermore, the dynamic rate is shown to be resilient to different variations in the network parameters without sacrificing packet latency.   
\end{abstract}


\acresetall		

\savespaces
\section{Introduction}\label{sec:intro}
 
{The \ac{IoT} is expected to extend wireless connectivity to a multitude of devices with sporadic traffic (e.g., measurements or updates)~\cite{GhaElsBadAlo:17, Chisci}, which necessitates traffic and topology aware performance characterization. In this regards, stochastic geometry and queueing theory are jointly utilized to account for the mutual interference and the underlying traffic requirements of the \ac{IoT} devices~\cite{GhaElsBadAlo:17,  Tony, Chisci, Fatma, ElSawy_deadline, AoI, Nardelli}. However, it is generally assumed that the generated packets can always fit within the time-slotted system regardless of the adopted transmission rate. However, rate adaptation require packet fragmentation to fit the time slot duration, which imposes a delicate tradeoff between reliability and latency.}

To the best of the author's knowledge, the tradeoff between rate adaptation and packet latency in slotted large-scale \ac{IoT} networks is still an open problem. This paper presents a novel spatiotemporal model to characterize such important tradeoff under static and dynamic rate adaptation schemes. Depending on the network parameters, we show that there exists a static rate that minimizes latency. In addition, we highlight the resilience of dynamic rate adaptation scheme, to different network parameters and packet sizes, at the expense of slightly increased latency.

\subsection{{Notations}}
{Upright and italic fonts are used, respectively, to denote random variables (e.g., $\rv{v}$) and their instantiations (e.g., $v$).  Vectors are bolded, e.g., $\V{v}$; matrices are bolded and uppercase, e.g., $\M{V}$. The notations $\M{I}$, $\M{0}$, and $\V{e}$ are used, receptively, for the identity matrix, the zeros matrix, and the column vector of ones, all with the appropriate sizes. The notation $[\cdot]^T$ denotes the transpose operator, $\mathcal{N}(\cdot)$ denotes the null space of a matrix, and $\M{V}^{[i,j]}$ denotes the $i^{\rm th}$-row $j^{\rm th}$-column element. $\M{V}^{[:,j]}$ is used to denote all elements in the $j^{\rm th}$ column within $\M{V}$. 
We use $\mathbb{P}\{\cdot\}$ and $\mathbb{E}\{\cdot\}$ to denote the probability and expectation, respectively. The over-bar denotes the complement operator, i.e., $\bar{v}\!=\!1\!-\!v$, $\|\cdot \|$ denotes the Euclidean norm, and $\mathbbm{1}_{\{\mathcal{E}\}}$ denotes the indicator function, which is equal to $1$ if the event $\mathcal{E}$ is satisfied and $0$ otherwise. }

\section{System Model}\label{sec:sytem}
\subsection{{Intended Link}} We consider an arbitrary IoT transmitter-receiver pair that are separated by $R_\circ$ meters. The transmitter emits $w_t$ Watts within the frequency band ($W$) to communicate to its intended receiver according to a time-slotted system with slot duration $T_s$. In each time slot, the intended transmitter generates a packet of size $L$ bits with probability $\alpha$. The intended traffic is parameterized with $L$ and $\alpha$ to reflect the quality and time resolutions of the generated measurements/updates. {Generated packets are stored in a buffer to be transmitted via a first in first out (FIFO) discipline. The transmitter can operate with one of $N$ different transmission rates $\mathcal{R}_1>\mathcal{R}_2>\cdots>\mathcal{R}_N$. The rate $\mathcal{R}_n$ is defined such that packets are fragmented to $n$ equal fragments, where the transmission of each fragment fits within one time slot duration $T_s$. Hence, $\mathcal{R}_n$ is given by}

{ \begin{equation} \label{eq:rate_dif}
 \mathcal{R}_n=\frac{L}{n \times T_s} =  \zeta W \log_2\left(1+\theta_n\right) ,
 \end{equation}
 where $0< \zeta \leq 1$ captures the gap between practical transmission rates when compared to Shannon's capacity\footnote{Due to practical considerations (e.g. finite blocklengh), it is only feasible to operate reliably at a certain percentage of the theoretical Shannon's capacity~\cite{Polyanskiy}.}} and  $\theta_n$ is the \ac{SIR} required to correctly decode the fragment at the receiver. Operating at  $\mathcal{R}_1$ (resp. $\mathcal{R}_N$) require at least $n=1$ (resp. $n=N$) time slots to deliver one packet. It is worth noting that $N <\frac{1}{\alpha}$ should be enforced for stable (i.e., finite latency) link operation. \\
\subsection{{Interference Model}} A heterogeneous Poisson field (HPF) of interferers is assumed.\footnote{The HPF implies an ad hoc network with  distributed administration where the intended and interfering links are deployed and operated by end users. Centralized administration  with unified parameters and operation is addressed in~\cite{Chisci, Fatma, ElSawy_deadline, Nardelli} for ad hoc networks and in \cite{GhaElsBadAlo:17, Tony,  AoI} for cellular networks.} The HPF is modeled via an arbitrary realization of a marked Poisson point process (PPP) $\mathrm{\Psi} \times \rv{v} \subset \mathbb{R}^2 \times \V{V}$. In particular, $\mathrm{\Psi} \subset \mathbb{R}^2$ is a PPP, with intensity $\lambda$, that denotes the locations of the interfering devices. Each interfering device is marked via a random mark $\rv{v}\in \V{V} = \{1, 2,\cdots, V\}$, which denotes the type of the IoT device.  Hence, the mark $\rv{v}\doteq v$ determines the transmission power $w_v \in \{w_1, w_2,\cdots, w_V\}$ and the activity factor $\kappa_v \in \{\kappa_1, \kappa_2,\cdots, \kappa_V\}$ of the IoT device. 
 The mark $\rv{v}$ has an arbitrary density function $f_{\rv{v}}(\cdot)$ that is independent from the devices locations. The HPF represents other types of coexisting IoT devices with locations and marks that are considered static once realized. Such assumption is justified by the short time slot duration that prohibits tangible variation in the locations or types of the coexisting devices. 
 \subsection{{Fading Model}} Transmitted signals are subject to both path-loss and multi-path Rayleigh fading. In particular, the signal power decays at the rate $r^{-\eta}$ with the distance $r$, where $\eta>2$ is the path-loss exponent. In addition, signals powers experience independent and identically distributed unit mean exponential block channel fading.  

\subsection{{Fragments Transmission Model}} {Due to fading and interference, transmission outages may occur. Particularity, when operating with rate $\mathcal{R}_n$, a fragment of size $\ceil{\frac{L}{n}}$ bits is successfully delivered to the intended receiver if the \ac{SIR} fulfills the decoding threshold $\theta_n= 2^{\frac{\mathcal{R}_n}{\zeta W}} -1 = 2^{\frac{L}{n \zeta W T_s}} -1.$}
Fragments that are correctly received at the intended receiver are dropped from the transmitter's buffer. Otherwise, retransmissions are attempted until successful packet delivery. 

We consider both static and dynamic rate adaptation schemes. In the static case, the rate $\mathcal{R}_n$ is predetermined offline and is never changed during the device operation. In the dynamic case, online rate control is realized via two parameters, namely the decrement probability $d$ and the increment probability $u$. Dynamic rate adaptation is only performed for the first fragment of each packet. Then, the same rate is utilized for all fragments within the same packet.\footnote{Fixing the transmission rate per packet reduces the overhead required for fragment tracking and decoding as well as packet reassembly at the receiver.} If the transmission of the first fragment at rate $\mathcal{R}_n$ fails, the intended transmitter opts to utilize $\mathcal{R}_{n+1}$ with probability $d$ and persists on $\mathcal{R}_n$ with probability $1-d$. Upon the successful transmission of the last fragment of a packet with rate $\mathcal{R}_n$, the intended transmitter explores $\mathcal{R}_{n-1}$ for the first fragment of the next packet with probability $u$ or exploits  $\mathcal{R}_{n}$ with probability $1-u$. {Such randomized rate adaptation scheme requires no control overhead and is adequate for IoT device-to-device links. }

  \section{Analysis}
By virtue of the independent thinning property of the PPP, we first split $\mathrm{\Psi}$ into $V$ independent PPPs denoted as $\mathrm{\Psi}_v$ with intensities $\lambda_v=f_{\rv{v}}(v) \lambda$. Let ${p}_n = \mathbb{P}\{\text{SIR} \geq \theta_n  \big| \mathrm{\Psi}\} $ be the probability that a fragment of size $\ceil{\frac{L}{n}}$ bits is successfully transmitted when operating at rate $\mathcal{R}_n$. {Without loss of generality, the intended receiver is assumed to be located at the origin. Accounting for the rate dependent SIR threshold,  $p_n$ can be expressed as

\begin{align}
p_n&\!=\! \mathbb{P}\left\{\frac{w_t \rv{h}_\circ \|\RV{x}_\circ\|^{-\eta}}{\sum_{v=1}^V \sum_{\RV{x}_i  \in \mathrm{\Psi}_v } \mathbbm{1}_{\{\mathcal{E}_{i}\}} w_v\; \rv{g}_i \; \|\RV{x}_i\|^{-\eta}}\! >\! \theta_n \; \big| \mathrm{\Psi} \right\} \notag \\
&{=} \prod_{v=1}^{V} \prod_{\RV{x}_i \in{\mathrm{\Psi}_v}} \left( {\frac{\kappa_v}{1+\theta_n \frac{ w_v R_\circ^{\eta} }{ w_t \|\RV{x}_i\|^{\eta} }}+ (1-\kappa_v)}\right),
\label{sinr1}
\end{align}
where $\RV{x}_\circ$ is the location of intended transmitter, $h_\circ$ is the intended channel gain, $\RV{x}_i \in \mathrm{\Psi}_v$ is the location of an interfering IoT device of type $v$, $g_i$ is the interfering channel gain, $\mathcal{E}_i$ is the event that the interfering IoT device is active. Note that \eqref{sinr1} is obtained substituting $R_\circ=\|\RV{x}_\circ\|$ and averaging over the devices types, activities, and fading gains. }

Due to the fixed realization of the HPF, the transmission success probability (TSP) is a function of the relative locations between the intended receiver and interfering IoT devices. Such realization dependent TSP is fully characterized via the meta distribution of the TSP~\cite{Hae:J16}, which is defined as
\begin{equation} \label{meta1}
\!\!\bar{F}_{s}(\theta_n,\gamma ) = \mathbb{P}\left\{ \mathbb{P}\left\{ \rm{SIR} > \theta_n \big| \mathrm{\Psi} \right\} > \gamma \right\} = \mathbb{P}\left\{ \rv{p}_n > \gamma \right\},
\end{equation} 
where the TSP ($\rv{p}_n$) is defined as a random variable to account for the different realizations of the HPF. In particular, for each rate $\mathcal{R}_n$, \eqref{meta1} defines the likelihood that the intended IoT link exists within a realization of the HPF such that the required decoding threshold $\theta_n$ is fulfilled for more than $\gamma$ percent of the time. Hence,  \eqref{meta1} generalizes the SIR model for the intended IoT link to all realizations of the HPF with a given set of parameters. For tractable analysis~\cite{Hae:J16}, the distribution in \eqref{meta1} is approximated as shown in the following proposition. 
\begin{proposition} \label{prop1}
 The likelihood that an IoT link successfully communicate with the transmission rate $\mathcal{R}_n$ for more than $\gamma$ percent of the time is approximated as

	\begin{align} \label{meta2}
	\bar{F}_{s}(\theta_n,\gamma)\! \approx \!  1\!-\! \mathcal{I}_\gamma \left(\frac{\mu_n (\mu_n- \nu_n)}{\nu_n- \mu_n^2},\frac{(1-\mu_n) (\mu_n-\nu_n)}{\nu_n-\mu^2} \right)\!\!,
	\end{align}
	\normalsize
	where $\mu_n$ and $\nu_n$ are the first two moments of the TSP at rate $\mathcal{R}_n$, and $\mathcal{I}_\gamma(a,b) = \frac{1}{\mathcal{B}(a,b)}\int_0^\gamma t^{a-1} (1-t)^{b-1} {\rm d}t$ is the regularized incomplete beta function. The moments  $\mu_n$ and $\nu_n$ are given by 
		\vspace{-0.4cm}
	\begin{align} \label{mom1}
\mu_n 	& =  \text{exp}\left\{ -\Upsilon \theta_n^\frac{2}{\eta}  \sum_{v=1}^V \left(\frac{w_v}{w_t}\right)^{\frac{2}{\eta}}\kappa_v \lambda_v \right\},
	\end{align}
	and
	\vspace{-0.4cm}
	\begin{align}  \label{mom2}
	\!\!\!\!\nu_n \!\!= \!\text{exp}\left\{\!- \Upsilon \theta_n^\frac{2}{\eta}  \sum_{v=1}^{V} \left(\frac{w_v}{w_t}\right)^{\frac{2}{\eta}} \kappa_v \lambda_v  \Big(\!2  \!-\! \Big(\!1-\!\frac{2}{\eta}\Big)\kappa_v \Big)   \right\}.
	\end{align}	
where $\Upsilon=\frac{2\pi^2 R_\circ^2}{\eta \sin (2\pi/\eta)}$.
\vspace{1mm}
	\begin{IEEEproof}
		The proof follows \cite{Hae:J16} while accounting for the rate dependent threshold $\theta_n$ and the HPF  parameterized by $V$, $f_\rv{v}(\cdot)$, $\V{w}$, and $\V{\kappa}$. 
	\end{IEEEproof}
\end{proposition}

To conduct the queueing analysis, we discretize the distribution in \eqref{meta2} for each rate $\mathcal{R}_n$ to $M$ equiprobable TSP classes. Each of the TSP classes captures all HPF realizations that would lead to a TSP within a certain range. To define the TSP ranges for the different classes, we set $\omega_{0}=0$ and $\omega_{M}=1$, and define the set  $\{\omega_{2},\omega_{3}, \cdots, \omega_{M-1}\}$ such that

	\begin{align} \label{meta3}
	\bar{F}_s(\theta_n,\omega_m)-\bar{F}_s(\theta_n,\omega_{m-1}) =\frac{1}{M}.
	\end{align}
	\vspace{-0.1cm}
	The TSPs within the range $[\omega_{m},\omega_{m+1}]$ are approximated via the median value $p_{n,m}$, which is given by

	\begin{align} \label{meta4}
	\bar{F}_{s}(\theta_n,\omega_m)\!-\!\bar{F}_{s}(\theta_n,p_{n,m}) \!=\! \frac{1}{2 M}.
	\end{align}
The above discritization implies that the likelihood for the intended IoT link to operate with any of the TSPs $p_{n,m}$ $\forall m$ is $\frac{1}{M}$. Using the discretized  $p_{n,m}$ $\forall (n,m)$ pairs, we can construct a queueing model for each TSP class for  static and dynamic rate adaptation schemes. Recall that each rate $\mathcal{R}_n$ spans the transmission of the packets over at least $n$ time slots. However, due to outages, the number of required time slots to deliver a packet is a random variable that varies across different TSP classes. For each TSP class, let $\rv{k}_{n,m}$ be a random variable representing the number of time slots required to deliver all fragments that belong to the same packet when operating with $\mathcal{R}_n$, then it is straightforward to show that $\rv{k}_{n,m}$ has the following distribution,    
\begin{align} \label{rate_dep}
\mathbb{P}\{\rv{k}_{n,m}=k | \mathcal{R}_n \}&=  \binom{k-1}{n-1} (p_{n,m})^n (1-p_{n,m})^{k-n},
\end{align}
where $n \leq k \leq \infty$ and $\mathbb{E}\left\{\rv{k}_{n,m}\right\}=\frac{n}{p_{n,m}}$. For each TSP class, the distribution in \eqref{rate_dep} captures the dependency between the transmission rate and latency for each packet. 

In the following, we construct the queueing model to track the buffer state at the intended transmitter while accounting for the TSP class and transmission rates. Utilizing the matrix analytic method (MAM)~\cite{alfa}, we use a hierarchical approach to construct the queueing transition matrix. At the top-level, $\M{P}$ tracks the number of packets in the transmitter's buffer. Since at most one packet can arrive or depart from the buffer at each time slot, the buffer states can be tracked via a quasi-birth-death (QBD) process with the following transition matrix

\begin{align}
\M{P}\!\!=\!\!\!\begin{bmatrix}
\M{B}  & \M{C}  & \M{0} & \M{0}& \M{0}  & \dots \\
\M{E} & \M{A}_1 &  \M{A}_0  & \M{0} & \M{0}  & \ddots \\
\M{0} & \M{A}_2  &\M{A}_1 &  \M{A}_0 & \M{0}   &\ddots \\
\vdots&  \ddots & \ddots &\ddots& \ddots & \ddots \\
\end{bmatrix},
\label{NB_MAT}
\end{align}
\normalsize
where the sub-matrices $\M{A}_2$, $\M{A}_1$, and $\M{A}_0$ track the transitions for which the number of packets in the buffer, respectively, decreases by one, remains constant, and increases by one. The sub-matrices $\M{B}$, $\M{C}$, and $\M{E}$ are the boundary matrices that track the transitions within, from, and to the idle state (i.e, empty buffer state). The detailed structure of the the sub-matrices within \eqref{NB_MAT} depends on the rate adaptation scheme (i.e., static or dynamic), the TSP class, and the transmission rate. 
\subsubsection{\textbf{Static Rate}} For a static rate of $\mathcal{R}_n$, we define the phase (PH) type distribution (see \cite[Sec. 2.5.3]{alfa}) with $1\times n$ initialization vector

\begin{equation} \label{beta_dif}
\V{\beta}_n= [1, 0, \cdots, 0],
\end{equation}
and $n\times n$ transient matrix

\begin{align}
\mathbf{T}_{n,m}\!\!=\!\!\!\begin{bmatrix} 
\bar{p}_{n,m} & p_{n,m}  & 0 &   \dots & 0 \\
0 & \bar{p}_{n,m} &  p_{n,m}  &\ddots & 0 \\
\vdots &    \ddots & \ddots &  \ddots & \vdots\\
0 &    \cdots &  0 & \bar{p}_{n,m} & p_{n,m} \\
0 &    \cdots & 0 & 0 &  \bar{p}_{n,m} \\
\end{bmatrix},
\label{T_static}
\end{align} 
\normalsize
which tracks the transmission success/failure of the $n$-fragments within each packet for each TSP class $m$ when operating with rate $\mathcal{R}_n$. Note that \eqref{beta_dif} and \eqref{T_static} are the PH-type representation of the distribution in \eqref{rate_dep}. Let $\V{s}_{n,m}=\V{e}-\mathbf{T}_{n,m}\V{e}$, where $\V{e}$ is a column vector of ones. Then the QBD for the static rate can be defined by the transition matrix in \eqref{NB_MAT} with the following parameters
\begin{equation}
\label{param_static}
\begin{matrix}
\text{$\M{B}= \bar{\alpha}$, $\;\;\M{C}=\alpha \V{\beta}_n$, $\;\;\M{E}=\bar{\alpha}  \M{s}_{n,m}$, $\;\M{A}_0 = {\alpha}  \M{T}_{n,m}$,} \\
\text{$\M{A}_1 = {\alpha}  \M{s}_{n,m} \V{\beta}_n  + \bar{\alpha}  \M{T}_{n,m}$, $\;\;\M{A}_2 = \bar{\alpha}  \M{s}_{n,m} \V{\beta}_n$}.
\end{matrix}
\end{equation} 
The stability of the static rate with such transition matrix is characterized in the following lemma,

\begin{lemma} \label{stab_static}
	For the TSP class $m\in\{1,2,\cdots,M\}$ and static transmission rate $\mathcal{R}_n$, the QBD model in \eqref{NB_MAT} with parameters in \eqref{param_static} is stable if and only if $\frac{p_{n,m}}{n} > \alpha$.    
%
	\begin{IEEEproof}
	{Follows from the distribution in \eqref{rate_dep}, where the system is stable if the packet departure rate $\frac{p_{n,m}}{n}$ is higher than the packet arrival rate $\alpha$. }
	\end{IEEEproof}
\end{lemma}
\begin{remark} {The stability condition in Lemma~\ref{stab_static} is important to determine the necessity of packet fragmentation. For a given traffic parameters $L$ and $\alpha$, the feasible rates $\mathcal{R}_n$ (i.e., that guarantees finite latency) should satisfy $\frac{n}{p_{n,m}} < \frac{1}{\alpha} $.}
\end{remark}
Let $\V{\pi}_{n,m}=\{{\pi}_{n,m,0},\V{\pi}_{n,m,1},\V{\pi}_{n,m,2},\cdots\}$ be the steady state solution of queueing model for the TSP class $m$ operating with static rate $\mathcal{R}_n$, where ${\pi}_{n,m,0}$ (resp. $\V{\pi}_{n,m,k}$) is the probability of having zero (resp. $k$) packets in the buffer. Note that  $\V{\pi}_{n,m,k}$ is a row vector representing 
the $n$ fragments that constitute the $k^{th}$ packet. For stable TSP classes, the steady state solution is characterized in the following theorem.

\begin{theorem} \label{sol_static}
	For the TSP class $m\in\{1,2,\cdots,M\}$ and static transmission rate $\mathcal{R}_n$ that satisfies Lemma~\ref{stab_static}, the steady state solution of the QBD shown in \eqref{NB_MAT} with parameters in \eqref{param_static} is given by
	
		  \begin{equation}
	\label{SS_sol_theorem1}
	\!\!\!\!\!\!\!	\V{\pi}_{n,m,k} \! =\!\left\{\begin{matrix}
	\left(1+\alpha\V{\beta}_n\M{Z}(\M{I}-\M{R}_{n,m})^{-1}\V{e}\right)^{-1} & \text{for} \; k=0 \\
	& \\
	{\pi}_{n,m,0}\alpha\V{\beta}_n\M{Z}& \text{for}\; k=1 \\
		& \\
	\V{\pi}_{n,m,1} \M{R}_{n,m}^{i-1} &  \text{for}\; k>1
	\end{matrix}\right., 	
	\end{equation}
	\normalsize
	where $\M{Z}=\left(\M{I}-\alpha\V{s}_{n,m} \V{\beta}_n - \bar{\alpha}\M{T}_{n,m}-\M{R}_{n,m} \bar{\alpha} \V{s}_{n,m} \V{\beta}_n\right)^{-1} $ and $\M{R}_{n,m}$ is the MAM rate matrix given by $\M{R}_{n,m} = \alpha \M{T}_{n,m} \left(\M{I}-{\alpha}  \M{s}_{n,m} \V{\beta}_n  - \bar{\alpha}  \M{T}_{n,m} - \alpha \M{T}_{n,m} \V{e} \V{\beta}_n\right)^{-1}$. 
	\begin{IEEEproof}
		Follows from the solution of Geo/PH/1 queueing systems \cite[Sec. 5.8]{alfa}, which is obtained by solving $\V{\pi}_{n,m} \; \M{P}= \V{\pi}_{n,m}$ along with the normalization condition $\V{\pi}_{n,m} \; \V{e} = 1$ for $\M{P}$ in \eqref{NB_MAT} with parameters in \eqref{param_static}. 
	\end{IEEEproof}	 
\end{theorem}

Let $\rv{z}_{n,m}$ be the number of time slots from packet generation to successful delivery, then the per-TSP average latency as well as the spatially averaged latency are given by
\begin{equation}
E\{\rv{z}_{n,m}\} \!=\! \frac{\V{\pi}_{n,m,1} (\M{I}\! -\!\M{R}_{n,m})^{-2} \V{e}}{a} \; 
\label{delay_stat}
\end{equation}
and 

\begin{equation}
E\{\rv{z}_{n}\}\! =\!\!\!\sum_{m=1}^{M} \frac{E\{\rv{z}_{n,m}\}}{M}
\label{delay_stat111}
\end{equation}

where $\V{\pi}_{n,m,1}$ and $\M{R}_{n,m}$ are  obtained from the steady state solution in Theorem~\ref{sol_static}. The latency in \eqref{delay_stat} follows from Little's law for discrete Markov chains, where ${\V{\pi}_{n,m,1} (\M{I} -\M{R}_{n,m})^{-2} \V{e}}$ is the average queue length. From the uniform discretization of the meta distribution, the spatially averaged latency in  \eqref{delay_stat111} (i.e., over all TSP classes) is obtained. 
\subsubsection{\textbf{Dynamic Rate}} 
We utilize a three-level hierarchy to construct the transition matrix for the dynamic rate. The top-level is defined by the QBD in \eqref{NB_MAT} to track the number of packets in the device's buffer. The middle-level of the hierarchy traces the rate adaptation and subsequent fragments transmissions through the matrices $\M{E}$ and $\M{C}$ along with 
\begin{equation}
\begin{matrix} \label{par_dynamic}
\text{$\M{B}=\bar{\alpha}\M{I}$, $\;\M{A}_0= \alpha \M{T}_m$, } \\ \text{$\;\M{A}_1= \bar{\alpha} \M{T}_m + {\alpha} \M{S}_m$, $\;\M{A}_2= \bar{\alpha} \M{S}_m$},
\end{matrix} 
\end{equation}
where $\M{T_m}$, $\M{S_m}$, $\M{E}$ and $\M{C}$ are shown in \eqref{S_T_Mat}-\eqref{F_E_C_Mat}. 
The matrix $\M{S}_m$ tracks the transmissions of the last fragment that belongs to a certain packet and the rate adaptation after successful packet transmission. The transmission of all other fragments are traced by the matrix $\M{T}_m$, which also tracks the rate adaptation for new packets. 

\begin{figure*}   
	\small
	\begin{align}
	\mathbf{T}_{m}\!\!=\!\!\!\begin{bmatrix}
	{0} &  \M{D}_{2,m} & \M{0}  & \dots & \M{0} \\
	\M{0}& \M{F}_{2,m} &   \M{D}_{3,m}  & \dots & \M{0} \\
	\vdots &\vdots & \ddots  & \ddots   & \vdots\\
	\M{0} &  \M{0}  &   \dots &  \M{F}_{N-1,m} & \M{D}_{N,m}\\
	\M{0} &  \M{0} & \dots &  \M{0} & \M{F}_{N,m}\\
	\end{bmatrix}, \;\; \text{\&} \;\;
	\mathbf{S}_m\!\!=\!\!\!\begin{bmatrix}
	p_{1,m} &  \M{0} & \M{0} &  \dots & \M{0} \\
	\M{U}_{1,m} & \M{M}_{2,m} &     \M{0} & \dots & \M{0} \\
	\M{0} &  \M{U}_{2,m}& \M{M}_{3,m}  &  \ddots &  \M{0} \\
	\vdots &\ddots & \ddots   & \ddots & \vdots\\
	\M{0}  &  \dots &  \M{0} &  \M{U}_{N-1,m} & \M{M}_N\\
	\end{bmatrix}.
	\label{S_T_Mat}
	\end{align} 
	\hrulefill 
	\begin{align}
	\mathbf{E}\!\!=\!\!\!\begin{bmatrix}
	p_{1,m} &  0 &  \dots & 0 \\
	\M{U}_{1,m}^{[:,1]}   & \M{M}_{2,m}^{[:,1]} &   \ddots & 0 \\
	\vdots  & \ddots  & \ddots  &  \vdots\\
	0  & \dots &    \M{U}_{N-1,m}^{[:,1]} &   \M{M}_{N,n}^{[:,1]}
	\end{bmatrix},	\;\; 
	\mathbf{C}\!\!=\!\!\!\begin{bmatrix}
	1  &   \M{0} &  \M{0} & \dots & \M{0} \\
	0 & \V{\beta}_{2} &  \M{0} & \dots & \M{0} \\
	\vdots  & \ddots  & \ddots  &  \ddots & \vdots\\
	0 &  \M{0} &     \dots &    \V{\beta}_{N-1}  &\M{0}\\
	0 &  \M{0} &   \dots &\M{0} &   \V{\beta}_{N} 
	\end{bmatrix}, \; \text{\&} \;
	\mathbf{F}_{n,m}\!\!=\!\!\!\begin{bmatrix}
	\bar{d} \; \bar{p}_{n,m} & p_{n,m}  &  0&  \dots & 0 \\
	0 & \bar{p}_{n,m} &  \ddots &\ddots & \vdots \\
	\vdots & \ddots &   \ddots &  \ddots & \vdots\\
	0 &    \cdots &  0 & \bar{p}_{n,m} & p_{n,m} \\
	0 &   \cdots & 0 & 0 &  \bar{p}_{n,m} \\
	\end{bmatrix}.	
	\label{F_E_C_Mat}
	\end{align}
	\hrulefill
	\begin{equation}
	\!\!\!\!\!\!\!\!\!\!\!\!	\M{D}_{n,m}^{[i,j]} =\left\{\begin{matrix}
	d \;  \bar{p}_{n-1,m}; &  \quad i=j=1 \\
	0; &   \text{otherwise}
	\end{matrix}\right., \quad
	\M{U}_{n,m}^{[i,j]} =\left\{\begin{matrix}
	u  \; p_{n,m}; &  \quad i=n \; \text{\&} \; j=1 \\
	0; &   \text{otherwise}
	\end{matrix}\right., 	\; \text{\&} \; 
	\M{M}_{n,m}^{[i,j]} =\left\{\begin{matrix}
	\bar{u}  \; p_{n,m}; &  \quad i=n \; \text{\&} \;  j=1 \\
	0; &   \text{otherwise}
	\end{matrix}\right.. 
	\label{D_U_M_Mat}	
	\end{equation}
	\hrulefill
	\normalsize
\end{figure*}
The matrices $\M{D}_{n,m}$, $\M{F}_{n,m}$, $\M{M}_{n,m}$, and $\M{U}_{n,m}$, shown in \eqref{F_E_C_Mat} and \eqref{D_U_M_Mat}, lie at the bottom of the hierarchy and represent the basic building blocks of the queueing system. In particular, $\M{D}_{n,m}$ is of size $(n-1)\times n$ and tracks the transition down from rate $\mathcal{R}_{n-1}$ to rate $\mathcal{R}_n$, the matrix $\M{F}_{n,m}$ is of size $n \times n$ and traces the fragments transmissions at rate $\mathcal{R}_n$ for one packet until the last fragment, the matrix $\M{M}_{n,m}$ is of size $n\times n$ and represents the transition after successfully transmitting the last fragment of a packet while persisting the same rate $\mathcal{R}_n$, and $\M{U}_{n,m}$ is of size $ (n+1)\times n$ and represents the transition after successfully transmitting the last fragment and going up from rate $\mathcal{R}_n$ to rate $\mathcal{R}_{n-1}$. Note that $\M{F}_{n,m}$ corresponds to $\M{T}_{n,m}$ in the static rate case that tracks the $n$ successful fragments transmissions required to deliver one packet at rate $\mathcal{R}_{n}$. 


Constructing $\M{T}_m$ and $\M{S}_m$  as shown in \eqref{S_T_Mat} with the elements in \eqref{F_E_C_Mat} and \eqref{D_U_M_Mat}, it can be shown that each has the size of $\Delta \times \Delta$, where $\Delta=\sum_{n=1}^{N} n =\frac{N (N+1)}{2}$. Note that the matrices $\M{B}$ (size $N\times N$), $\M{E}$ (size $\Delta \times N$), and $\M{C}$ (size $N \times \Delta$) are defined to keep track of the last utilized rate in transmissions before the buffer becomes empty. Hence, empty buffers do not interrupt the memory of the employed rate adaptation scheme. Once the transition matrix $\V{P}$ is constructed, the next step is to check the stability of the queueing model for each TSP class.

   	\begin{figure}[t]
   \begin{tikzpicture}[scale=0.6, transform shape,font=\Large]
   			\input{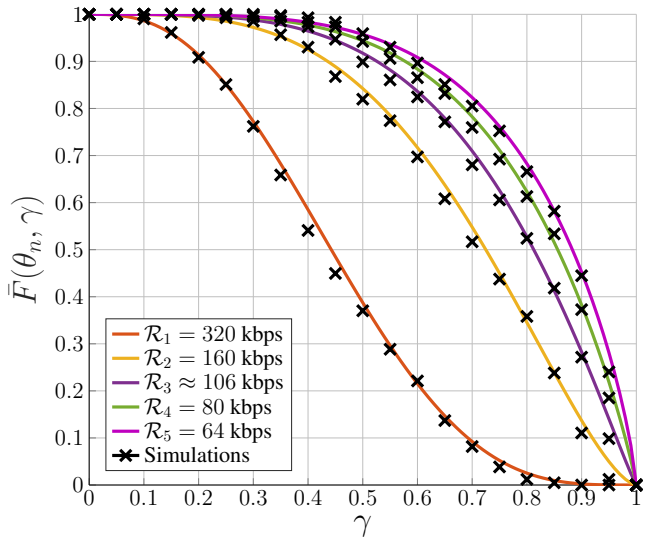}
   			\end{tikzpicture}
   		   		   		\caption{Meta distribution of the TSP for $w_t=10$ mW.}
   		\label{fig:meta}
   	\end{figure}

   	\begin{figure}[t]
   	\begin{tikzpicture}[scale=0.6, transform shape,font=\Large]
%
%
\definecolor{mycolor1}{rgb}{0.00000,0.44700,0.74100}%
\definecolor{mycolor2}{rgb}{0.85000,0.32500,0.09800}%
\definecolor{mycolor3}{rgb}{0.92900,0.69400,0.12500}%
\definecolor{mycolor4}{rgb}{0.49400,0.18400,0.55600}%
\definecolor{mycolor5}{rgb}{0.46600,0.67400,0.18800}%

\begin{axis}[%
width=5.18in,
height=1.333in,
at={(0.869in,0.487in)},
scale only axis,
bar shift auto,
xmin=0.5,
xmax=8.5,
xtick={1, 2, 3, 4, 5, 6, 7, 8},
xlabel style={font=\color{white!15!black}},
xlabel={\huge{TSP class}},
ymin=0,
ymax=40,
ylabel style={font=\color{white!15!black}},
ylabel={\huge{Total Latency}},
axis background/.style={fill=white},
ymajorgrids,
legend style={legend cell align=left, align=left, draw=white!15!black}
]
\addplot[ybar, bar width=0.145, fill=mycolor1, draw=black, area legend] table[row sep=crcr] {%
1	11.4834080998924\\
2	7.27648057525873\\
3	5.90248804482811\\
4	5.11774739118408\\
5	4.57096386586419\\
6	4.14153043118767\\
7	3.76518199278899\\
8	3.35786290973636\\
};
\addplot[forget plot, color=white!15!black] table[row sep=crcr] {%
0.5	0\\
8.5	0\\
};

\addplot[ybar, bar width=0.145, fill=mycolor2, draw=black, area legend] table[row sep=crcr] {%
1	inf\\
2	inf\\
3	83.3612348036373\\
4	31.8734616496944\\
5	18.7061584330594\\
6	12.3976968700302\\
7	8.43416721974241\\
8	5.23098604564351\\
};
\addplot[forget plot, color=white!15!black] table[row sep=crcr] {%
0.5	0\\
8.5	0\\
};

\addplot[ybar, bar width=0.145, fill=mycolor3, draw=black, area legend] table[row sep=crcr] {%
1	12.3126671384278\\
2	7.07067460722761\\
3	5.46293176913482\\
4	4.56414844962133\\
5	3.94414042708189\\
6	3.45929285720973\\
7	3.03570711334324\\
8	2.59025014600853\\
};
\addplot[forget plot, color=white!15!black] table[row sep=crcr] {%
0.5	0\\
8.5	0\\
};

\addplot[ybar, bar width=0.145, fill=mycolor4, draw=black, area legend] table[row sep=crcr] {%
1	11.4276658664518\\
2	7.25664486396153\\
3	5.89388915600803\\
4	5.11713454912359\\
5	4.57838110594218\\
6	4.15929190752119\\
7	3.8003703588445\\
8	3.44409121262942\\
};
\addplot[forget plot, color=white!15!black] table[row sep=crcr] {%
0.5	0\\
8.5	0\\
};

\end{axis}

\begin{axis}[%
	width=5.18in,
	height=1.333in,
	at={(0.869in,2.398in)},
	scale only axis,
	bar shift auto,
	xmin=0.5,
	xmax=8.5,
	xtick={1, 2, 3, 4, 5, 6, 7, 8},
	xlabel style={font=\color{white!15!black}},
	xlabel={\huge{TSP class}},
	ymin=0,
	ymax=27,
	ylabel style={font=\color{white!15!black}},
	ylabel={\huge{Tx Latency}},
	axis background/.style={fill=white},
	ymajorgrids,
    legend style={at={(0.19,1.05)}, anchor=south west, legend columns=5, legend cell align=left, align=left, draw=white!15!black}
	]
	\addplot[ybar, bar width=0.145, fill=mycolor1, draw=black, area legend] table[row sep=crcr] {%
		1	8.96567688214733\\
		2	6.31419695649225\\
		3	5.30433569897858\\
		4	4.69194045906171\\
		5	4.24915680275136\\
		6	3.89173799698488\\
		7	3.57103392508677\\
		8	3.21440836461923\\
	};
	\addplot[forget plot, color=white!15!black] table[row sep=crcr] {%
		0.5	0\\
		8.5	0\\
	};
	\addlegendentry{\Large{Dyn.}}
	
	\addplot[ybar, bar width=0.145, fill=mycolor2, draw=black, area legend] table[row sep=crcr] {%
		1	65.5157736917952\\
		2	29.4358200565164\\
		3	19.4113906560651\\
		4	14.2614493127028\\
		5	10.9505040487198\\
		6	8.51544049222414\\
		7	6.50098949866717\\
		8	4.4738364602852\\
	};
	\addplot[forget plot, color=white!15!black] table[row sep=crcr] {%
		0.5	0\\
		8.5	0\\
	};
	\addlegendentry{\Large{$\mathcal{R}_1$}}
	
	\addplot[ybar, bar width=0.145, fill=mycolor3, draw=black, area legend] table[row sep=crcr] {%
		1	9.03938877800943\\
		2	5.97659317056135\\
		3	4.83368095319277\\
		4	4.14502159801307\\
		5	3.64762720084795\\
		6	3.24536468891086\\
		7	2.88404151485853\\
		8	2.49380976374945\\
	};
	\addplot[forget plot, color=white!15!black] table[row sep=crcr] {%
		0.5	0\\
		8.5	0\\
	};
	\addlegendentry{ \Large{$\mathcal{R}_2$}}
	
	\addplot[ybar, bar width=0.145, fill=mycolor4, draw=black, area legend] table[row sep=crcr] {%
		1	8.79125692105585\\
		2	6.21616114743993\\
		3	5.23453772869604\\
		4	4.64005351848613\\
		5	4.21204014030134\\
		6	3.86992720704241\\
		7	3.5704027023977\\
		8	3.26700780983865\\
	};
	\addplot[forget plot, color=white!15!black] table[row sep=crcr] {%
		0.5	0\\
		8.5	0\\
	};
	\addlegendentry{ \Large{$\mathcal{R}_3$}}
	
	\addplot [color=black, dashed, line width=2.0pt]
	table[row sep=crcr]{%
		0	25\\
		1	25\\
		2	25\\
		3	25\\
		4	25\\
		5	25\\
		6	25\\
		7	25\\
		8	25\\
		9	25\\
	};
	\addlegendentry{\Large{Thres.}}
	
\end{axis}
   			\end{tikzpicture}
   		   		\caption{Tx and queueing latency for $w_t=50$ mW, where infinite queueing latencies for unstable TSP classes are omitted from the bottom figure.  The legend (Dyn.) stands for dynamic rate and (Thres. = $\frac{1}{\alpha})$) stands for stability threshold.  }
   		\label{fig:delay1}
   	\end{figure}
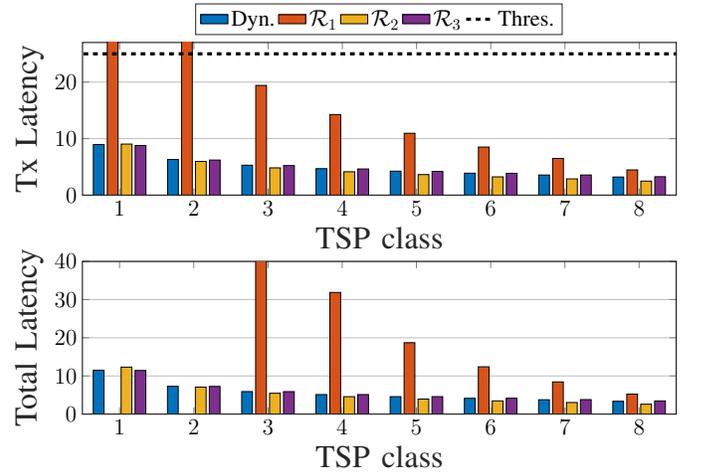 
   	
   	\begin{figure}[t]
   \begin{tikzpicture}[scale=0.6, transform shape,font=\Large]
   			\input{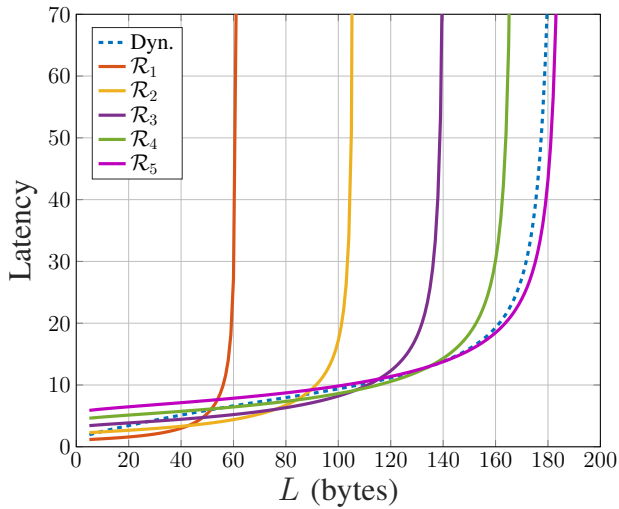}
   			\end{tikzpicture}
   		   		\caption{Average latency for $w_t=50$ mW where $\mathcal{R}_n\!=\!\frac{8L}{n}$kbps.  The legend (Dyn.) stands for dynamic rate.}
   		\label{fig:delay2}
   	\end{figure} 

  \begin{lemma} \label{stab_dyn}
  	For the TSP class $m\in\{1,2,\cdots,M\}$, the QBD model in \eqref{NB_MAT} with the parameters in \eqref{par_dynamic}-\eqref{D_U_M_Mat} is stable if and only if $\bar{\alpha}\V{\Pi}_m \M{S}_m\V{e} > {\alpha}\V{\Pi}_m  \M{T}_m  \V{e}$, where ${\V{\Pi}_m=\Big(\big[\M{S}_m+\M{T}_m\big]^T-\M{I}+\V{e}\V{e}^T\Big)^{-1}\V{e}}$. 
  	\begin{IEEEproof}
  		Stability requires $\V{\Pi}_m \M{A}_2 \V{e} > \V{\Pi}_m  \M{A}_1 \V{e}$~\cite{alfa}. The vector $\V{\Pi}_m$ is the solution of $\V{\Pi}_m \M{A} = \V{\Pi}_m$ and $\V{\Pi}_m \V{e} = 1$,
  		where $\M{A} =\M{A}_0 + \M{A}_1 + \M{A}_2 $. The lemma follows by applying the specific structure of $\M{A}_0$, $\M{A}_1$, and $\M{A}_2$. Note that the explicit expression for $\V{\Pi}_{m}$ is obtained by following~\cite{Krikidis}. 
  		\end{IEEEproof}
  \end{lemma}
  For stable TSP classes, the steady state solution for the dynamic rate scheme is characterized in the following theorem
  	 \begin{theorem} \label{them2}
  	 For the TSP class $m\in\{1,2,\cdots,M\}$ that satisfies Lemma~\ref{stab_dyn}, the steady state solution of the QBD shown in \eqref{NB_MAT} with the parameters in \eqref{par_dynamic}-\eqref{D_U_M_Mat} is given by
  	
  	 	   \begin{equation}
  	 	   \label{SS_sol_theorem}
  	 	\!\!\!\!\!\!\!\!\!\!\!\!	\V{\pi}_{m,k} \! =\!\left\{\begin{matrix}
  	 	\frac{(1-a)\V{\pi}_{m,1} \V{E}_m}{a} & \text{for} \; k=0 \\
  	 	& \\
  	 	\mathcal{N}\Big(\M{Q}_m \Big)& \text{for}\; k=1 \\
  	 	& \\
  	 	\V{\pi}_{m,1} \M{R}_m^{i-1} &  \text{for}\; k>1
  	 	\end{matrix}\right., 	
  	 	\end{equation}
  	 	\normalsize
  	 	such that  $\M{Q}_m = \big[(1-a)(\M{E}_m \M{C}+\M{T}_m +\M{R}_m \M{S}_m)+a\M{S}_m\big]^{T}-\M{I}$ and the  condition $\V{\pi}_{m,0} \V{e}+ \V{\pi}_{m,1} (\M{I}-\M{R}_m)^{-1} \V{e}=1$ is satisfied, where $\mathcal{N}(\cdot)$ denotes the null space of a matrix and  $\M{R}_{m}$ is the MAM rate matrix obtained by the minimum non-negative solution of $\M{R}_{m} = \alpha \M{T}_{m} + \M{R}_{m} \left( \alpha \M{S}_{m} + \bar{\alpha} \M{T}_{m} \right)+ \bar{\alpha} \M{R}_{m}^2  \M{S}_{m}$.
  	 \begin{IEEEproof}
  	 	Follows from the steady state solution of QBDs~\cite[Sec. 4.8]{alfa}, which is obtained by solving $\V{\pi}_{m} \; \M{P}= \V{\pi}_{m}$ along with the normalization condition $\V{\pi}_{n,m} \; \V{e} = 1$ for $\M{P}$ in \eqref{NB_MAT} with parameters in \eqref{par_dynamic}-\eqref{D_U_M_Mat}. Note that there is no explicit closed-form expression for $\M{R}_{m}$ since neither $\M{A}_2$ nor $\M{A}_0$ are of rank 1~\cite[Sec. 4.8]{alfa}. However, there are efficient algorithms to find $\M{R}_m$ as shown in~\cite{alfa, Fatma, QBD_solve}.
  	 	\end{IEEEproof}	 
  	 \end{theorem}
  Similar to \eqref{delay_stat} and \eqref{delay_stat111}, utilizing $\V{\pi}_{m,1}$ and $\M{R}_{m}$ given in Theorem~\ref{them2}, the per TSP class latency and spatially averaged latency for the dynamic rate are given by

   \begin{align}\label{delay_dyn}
   E\{\rv{z}_{m}\} &= \frac{\V{\pi}_{m,1} (\M{I} -\M{R}_{m})^{-2} \V{e}}{a}, 
   \end{align} 

and 

 \begin{align}\label{delay_dyn111}
   E\{\rv{z}\} = \sum_{m=1}^{M} \frac{E\{\rv{z}_{m}\}}{M}. 
\end{align}

%
\section{Numerical Results}\label{sec:conclude}
 { Unless otherwise stated, the HPF parameters are $\lambda=10^3$ device/km$^2$, that are divided into $V=3$ networks, with uniform probability distribution $f_{\rv{v}}(v)=1/3$, activity factors $\V{\kappa}=\{0.1, 0.3, 0.5\}$, and powers $\V{w}=\{10, 7, 5\}$ mWatts. The intended link parameters are $\alpha=0.04$, $W=100$ kHz, $\zeta=0.8$, $T_s= 1$ms, $\eta=4$, $R_\circ=20$ meters, $N=5$ rates, $M=8$ TSP classes, $L=40$ bytes, $d=0.3$, and $u=0.1$. Hence, $\mathcal{R}_n= \frac{320}{n}$ kbps and $\theta_n=2^{4/n}-1$. The Monte Carlo simulations construct $\bar{F}_{sim}(\theta_n,\gamma)$ across 5000 different HPF realizations with $10^5$ time iteration per each HPF realization.}
  
  {For the analysis, the meta distribution of the TSP for each rate $\mathcal{R}_n$ is obtained via  \eqref{meta2}. To compute latency, the TSP classes $p_{n,m}$, $\forall (n,m)$ pairs are obtained from \eqref{meta4}, which discretize the meta distribution in \eqref{meta2}. The TSP classes are then used to populate the sub-matrices in \eqref{NB_MAT}, which are given in \eqref{T_static}-\eqref{param_static} (resp. \eqref{par_dynamic}-\eqref{D_U_M_Mat}) for static (resp. dynamic) rate. The stability of the queueing model can be verified by Lemma~\ref{stab_static} (resp. Lemma~\ref{stab_dyn}) for static (resp, dynamic) rate. For stable queues, the steady state solution is obtained from \eqref{SS_sol_theorem1} (resp. \eqref{SS_sol_theorem}) for static (resp. dynamic) rate. Once the steady state solution is determined, the delay can be computed via \eqref{delay_stat} (resp. \eqref{delay_dyn}) for static (resp. dynamic) rate. }
  
Fig.~\ref{fig:meta} shows the meta distribution of the TSP for different transmission rates, where the close match between the analysis (i.e., curves) and simulations (marks) validates Proposition~\ref{prop1}. The figure also shows the impact of $\mathcal{R}_n$ on the TSP (i.e., transmission reliability). Dividing the packet into more fragments enables a lower transmission rate. Hence, leading to a lower detection threshold  $\theta_n$ that is more likely to be satisfied, which improves transmission reliability. However, at the expense of expanding the packet transmission over multiple time slots.

Fig.~\ref{fig:delay1} shows the latency for different transmission rates. The top (resp. bottom) figures of Fig.~\ref{fig:delay1} show the transmission (resp. overall) latency for different TSP classes. The figure shows that each TSP class has its own adequate transmission rate. Importantly, for poor TSP classes fragmenting the packet over multiple time slots is necessary for stable operation. {For instance, operating at $\mathcal{R}_1$ for TSP classes 1 and 2 leads to a transmission latency that exceeds the stability threshold (i.e., $\alpha^{-1}$), and hence, the latency is infinite (note that infinite latencies are omitted from the bottom figure of  Fig.~\ref{fig:delay1}).} The figure also shows that there is an optimal static rate for different TSP classes, and hence, packet fragmentation depends on the network parameters and the HPF realization. The figure also shows that the dynamic rate can achieve low latency across all TSP classes, which confirms its resilience. 

Fig.~\ref{fig:delay2} shows the spatially averaged (i.e., over all TSP classes) latency vs the packet size for different transmissions rates. The figure shows that the transmission rate and subsequent fragmentation depend on the underlying packet size. In compliance with intuition, larger packets should be divided into more fragments in order to operate with a reliable transmission rate. {For instance, packets from 45 to 75 bytes  (resp. 75 to 110 bytes) should be divided into two (resp. three) fragments to achieve minimal delay. The figure also shows that over-fragmentation of packets lead to unnecessary delay (e.g., $L<45$ bytes).} Interestingly, the figure shows a resilient performance for the dynamic rate scheme at the expense of a slightly prolonged latency. Note that the latency degradation of the dynamic scheme is due to its  randomized exploration of different transmission rates.      

\savespaces
\section{Conclusion}\label{sec:conclude}
Using tools from stochastic geometry and queueing theory, this paper develops a novel mathematical model for static and dynamic rate adaptation schemes in IoT networks. The developed model accounts for the effect of transmission rate adaptation on packet fragmentation to comply with the time slot duration, which captures an important tradoff between transmission reliability and packet latency. It is shown that fragmentation is sometimes necessary to maintain finite packet latency. Also, over-fragmentation of packets may lead to unnecessary delays. Hence, there exists an optimal fragmentation (i.e., static rate) that minimizes packet latency, which depends on the network parameters. {To this end, the paper reveals the resilience of a simple randomized dynamic rate adaptation to different variations in the packet sizes and network parameters without sacrificing the latency performance.   }



\savespaces
\bibliographystyle{IEEEtran}
\bibliography{IEEE-conf-abrv,StringDefinitions,Ref}

\end{document}